\documentclass[ aps,%
floatfix,%
final,%
notitlepage,%
oneside,%
onecolumn,%
nobibnotes,%
nofootinbib,%
superscriptaddress,%
showpacs,%
]%
{revtex4}

\usepackage{epsfig}
\usepackage{amsfonts}
\usepackage{amsmath}
\usepackage{epsfig}
\usepackage{graphics}

\newcommand{\JP}{\psi}

\newcommand{\beq}{\begin{eqnarray}}\newcommand{\eeq}{\end{eqnarray}}
\newcommand{\beqa}{\begin{eqnarray*}}\newcommand{\eeqa}{\end{eqnarray*}}

\begin{document}

\title{ The study of leading twist light cone wave functions of $2S$ state charmonium mesons.}
\author{V.V. Braguta}
\email{braguta@mail.ru}
\affiliation{Institute for High Energy Physics, Protvino, Russia}

\begin{abstract}
In this paper leading twist light cone wave functions of $2S$ state charmonium mesons are studied 
and models of these functions are built. 
\end{abstract}

\maketitle

\newcommand{\ins}[1]{\underline{#1}}
\newcommand{\subs}[2]{\underline{#2}}
\vspace*{-1.cm}
\section{Introduction.}

Charmonium light cone wave functions (LCWF) are universal nonperturbative objects that describe
the production of charmonium mesons in hard exclusive processes within light cone formalism \cite{Chernyak:1983ej}. 
Usually to study hard exclusive processes with quarkonium production one uses NRQCD \cite{Bodwin:1994jh}. 
So, light cone formalism can be considered as alternative to NRQCD. 

There are two very important advantages of light cone formalism in comparison to NRQCD. 
The first one is connected with the following fact: light cone formalism can be applied to study 
the production of any meson. For instance, it is possible to study the production of light mesons, 
such as $\pi$ mesons, or the production of heavy mesons, such as charmonium mesons, if LCWFs of these 
mesons are known. From the NRQCD perspective, this implies that light cone formalism resums infinite 
series of relativistic corrections to the amplitude, what can be very important 
\cite{Braaten:2002fi, Liu:2002wq, Liu:2004ga, Zhang:2005ch, Bodwin:2006ke, He:2007te}. The second advantage is that light cone 
formalism easily resums leading logarithmic radiative corrections to the amplitude $\sim \alpha_s \log (Q)$
with the help of LCWFs. This is very important advantage since leading logarithmic corrections at high energies 
can be even more important than relativistic corrections to the amplitude. 

From this one can conclude that LCWFs are the key ingredient of light cone formalism. Moreover, the 
universality of LCWFs and the variety 
of the processes where these functions can be used make the study of charmonium LCWFs to be a very important 
task. However, despite the fact that charmonium LCWFs are very important in understanding hard exclusive 
processes with charmonium production there is a very limited knowledge of the properties of these functions.
There are only few papers where this functions were studied \cite{Bodwin:2006dm, Ebert:2006xq, Ma:2006hc, Choi:2007ze}.

In this paper the procedure developed in papers \cite{Braguta:2006wr, Braguta:2007fh} for the study 
of charmonium LCWFs will be applied to the study of leading twist LCWFs of $\Psi'$ and  $\eta_c'$ mesons. 
This paper is organized as follows. In the next section all definitions needed in the calculation will be given. 
In Section III the moments of LCWFs will be calculated in the framework of Buchmuller-Tye and Cornell potential models. 
Section IV is devoted to the calculation of the moments within NRQCD. QCD sum rules will be applied 
to the calculation of the moments in Section V. Using the results obtained in Sections III-V 
the models of LCWFs will be built in Section VI. In the last section the results of this paper will be summarized.

\section{ Definitions.}
There is one leading twist light cone wave function (LCWF) of $\eta_c'$ meson $\phi_{\eta} (\xi, \mu)$
 and there are two leading twist LCWFs of $\JP'$ meson $\phi_L ( \xi, \mu),~ \phi_T ( \xi, \mu)$. 
The function $\phi_L ( \xi, \mu)$ is twist two LCWF of longitudinally polarized $\JP'$ meson. The function $\phi_T ( \xi, \mu)$ is 
twist two LCWF of transversely polarized $\JP'$ meson.
These LCWFs can be defined as follows \cite{Chernyak:1983ej}
\beq
\nonumber
{\langle 0| {\bar Q} (z) \gamma_{\alpha} \gamma_5 [z,-z] Q(-z) | \eta_c'(p) \rangle}_{\mu} &=& i
f_{\eta} p_{\alpha} \int^1_{-1} d \xi \,e^{i(pz) \xi}
\phi_{\eta} ( \xi, \mu), \\
{\langle 0| {\bar Q} (z) \gamma_{\alpha}  [z,-z] Q(-z) | \JP'(\epsilon_{\lambda=0} ,p) \rangle}_{\mu} &=&
f_{L} p_{\alpha} \int^1_{-1} d \xi \,e^{i(pz) \xi}
\phi_L ( \xi, \mu), \nonumber \\
{\langle 0| {\bar Q} (z) \sigma_{\alpha \beta}  [z,-z] Q(-z) | \JP'(\epsilon_{\lambda=\pm 1} ,p) \rangle}_{\mu} &=&
f_{T} (\mu) ( \epsilon_{\alpha} p_{\beta} - \epsilon_{\beta} p_{\alpha} )  \int^1_{-1} d \xi \,e^{i(pz) \xi}
\phi_T ( \xi, \mu),
\label{lwf}
\eeq
where the following designations are used: $x_1, x_2$ are the momentum fractions of the whole meson carried by quark and antiquark 
correspondingly, $\xi = x_1 - x_2$, $p$ is the momentum of corresponding meson, $\mu$ is an energy scale. The factor $[z,-z]$, 
makes the matrix elements  to be gauge invariant and the dependence of the LCWFs $\phi_{0, L, T} (x, \mu)$ on scale 
$\mu$ can be found in \cite{Chernyak:1983ej, Braguta:2006wr, Braguta:2007fh}. 

It should be noted here that there is an important distinction between LCWF of $\eta_c'$ and $\psi'$ mesons. 
Let us, for instance, consider LCWF of longitudinally polarized $\JP'$. Obviously, this function can be written 
as follows
\beq
\phi_L ( \xi, \mu)=\phi_L^S ( \xi, \mu) + \phi_L^D ( \xi, \mu),
\eeq
where $\phi_L^S ( \xi, \mu)$ and $\phi_L^D ( \xi, \mu)$ are $S$- and $D$-wave contributions to 
LCWF of $\psi'$ meson. In the case of $\eta_c'$ meson only $S$-wave contributes to the LCWF of this meson. 
It is not difficult to estimate the contribution of $D$-wave to LCWF $\phi_L ( \xi, \mu)$. Evidently, $D$-wave 
admixture in the LCWF $\phi_L ( \xi, \mu)$ is proportional to the factor $\sim \tan (\theta) f_L^D/f_L^S$, where 
$f_L^S, f_L^D$ are $S$- and $D$-wave contributions to the constant $f_L$, $\theta$ is a mixing
angle of $S$ and $D$ waves in $\psi'$ meson. Within potential models this factor can be written as 
\beq
\tan (\theta) f_L^D/f_L^S = \tan (\theta) \frac 5 {\sqrt{8} M_c^2} \frac {R_D''(0)} {R_S(0)} 
\eeq
where $M_c$ is a quark mass in the framework of potential model, $R_S(r), R_D(r)$ are radial wave 
function of $D$ and $S$ waves. Numerical values of parameters $\theta, R_D''(0), R_S(0), M_c$ 
needed for the estimation of $D$-wave contribution to LCWF of $\psi'$ meson will be taken from 
paper \cite{Rosner:2001nm}: $R_S(0)=0.734$~GeV$^{3/2}$, $5 R_D''(0) / \sqrt{8} M_c^2=0.095$~GeV$^{3/2}$, $\theta \sim 12^o$.
Thus one gets rather large suppression of $D$-wave admixture $\tan (\theta) f_L^D/f_L^S \sim 0.03$. 
On account of the considerable suppression $D$-wave admixture one can disregard its contribution 
to the LCWFs of $\psi'$ meson. Below this approximation will be used.  

Commonly, $\eta_c'$ and $\psi'$ mesons are considered as a nonrelativistic bound states of 
quark-antiquark pair. At leading order approximation in relative velocity of 
quark-antiquark pair $\eta_c'$ and $\psi'$ mesons cannot be distinguished. So within this 
approximation $\eta_c'$ and $\psi'$ mesons have identical LCWFs at scale $\mu \sim M_c$
\beq
\phi_{\eta} ( \xi, \mu) =  \phi_{L} ( \xi, \mu) =  \phi_{T} ( \xi, \mu) = \phi (\xi, \mu).
\label{hy}
\eeq
One can expect that in the case of $2S$ mesons corrections to this approximation 
can be large. However, the accuracy obtained in this paper does not allow one to 
distinguish LCWFs $\phi_{\eta, L, T} (x, \mu)$. For this reason approximation 
(\ref{hy}) will be used in this paper.

The main goal of this paper is to calculate the LCWFs $\phi_{\eta, L, T}(\xi, \mu)$ of $\JP'$ and $\eta_c'$ mesons.
These LCWFs will be parameterized by their moments $\langle \xi^n_{\eta, L, T}  \rangle_{\mu}$ at some scale.
It is worth noting that, the LCWFs (\ref{lwf}) are $\xi$-even, so 
only even moments should be calculated.

\section{ The moments in the framework of potential models. }

\begin{figure}[t]
\begin{center}
\includegraphics[scale=0.8]{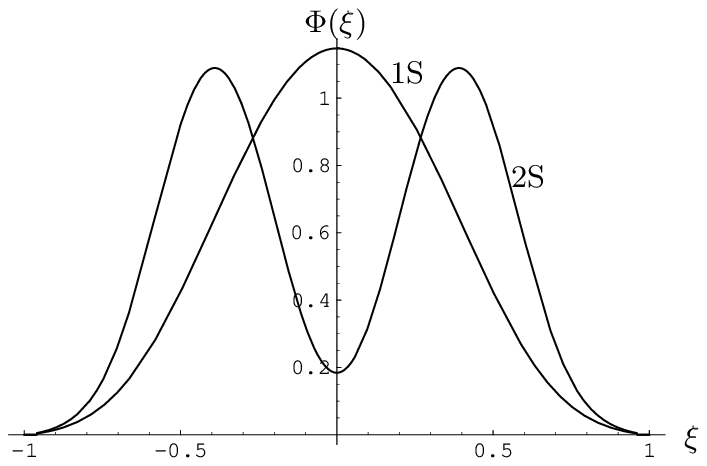} 
\caption{The functions $\Phi( \xi)$ for 1$S$ and 2$S$ states.}
\end{center}
\end{figure}

In papers \cite{Braguta:2006wr, Braguta:2007fh} it was shown that the moments of LCWFs of 
$\eta_c$ and $J/ \JP$ mesons  can be calculated in the framework of potential models. 
In comparison with QCD sum rules, such calculation cannot be considered as an accurate one. 
However, potential models give rather good estimation of the values of the moments. 

To calculate the moment of LCWF one can apply Brodsky-Huang-Lepage (BHL) \cite{Brodsky:1981jv} procedure
that can be written as 
\beq
\nonumber
\phi (\xi, \mu) &\sim& ~ \phi_{as} (\xi) \Phi( \xi, \mu)= (1-\xi^2) \Phi( \xi, \mu), \\
\Phi(\xi, \mu) &=& \int_0^{ \frac {\mu^2} {1- \xi^2} } d t ~ \psi \biggl ( t + \frac {\xi^2 M_c^2} {1-\xi^2} \biggr ),
\label{p2}
\eeq
where $\psi ( { \bf k^2})$ is the solution of Schrodinger equation in momentum space, $M_c$ is a 
quark mass within potential model. 

In this paper the function $\psi( {\bf k^2})$ will be calculated in the framework of the potential models with 
Buchmuller-Tye \cite{Buchmuller:1980su} and Cornell potentials \cite{Eichten:1978tg}. The parameters of
Buchmuller-Tye potential model will be taken from paper \cite{Buchmuller:1980su}. For Cornell potential $V(r) = -k/r+ r/a^2$
the calculation will be carried out with the following set of parameters: 
$k=0.358,~ a=2.381 \mbox{~GeV}^{-1},~ M_c=1.147$ GeV \cite{Bodwin:2006dm}. The scale 
$\mu$ is taken equal to $1.5$ GeV. 

The results of our calculation are presented in Table I. In second and third columns the moments calculated in the framework of 
Buchmuller-Tye and Cornell models are presented. It is seen that there is  good agreement between these two 
models. In papers \cite{Braguta:2006wr, Braguta:2007fh} it was shown 
that potential models cannot be applied for higher moments. Due to this fact the calculations have 
been restricted by few first moments.

Now the following point deserves consideration. It was shown in equations (\ref{p2}) that LCWFs can be represented 
as a product of the asymptotic function $\phi_{as} (\xi)$ and the function $\Phi (\xi, \mu)$. The function 
$\Phi(\xi, \mu)$ contains information about the internal motion of quark antiquark pair in meson. Let us compare these functions 
$\Phi( \xi)$ for 1$S$ and 2$S$ states. These functions calculated within Buchmuller-Tye potential model and 
normalized as $\int d \xi \Phi ( \xi) = 1$ are shown in Fig. 1. It is seen that the function $\Phi( \xi)$ of 1$S$ state
has rather simple shape with one extremum at $\xi=0$. It is not difficult to guess that this extremum appears since the 
function $\Phi (\xi)$ is $\xi-$even. For the same reason the function $\Phi(\xi)$ of 2$S$ state has similar extremum
at $\xi=0$. However, the function $\Phi(\xi)$ of 2$S$ state has two additional extremums located symmetrically relative 
to $\xi-$axis. 

\begin{table}
$$\begin{array}{|c|c|c|c|c|}
\hline \langle \xi^n \rangle & \mbox{ Buchmuller-Tye } & \mbox{ Cornell } & \mbox{  NRQCD }
& \mbox {QCD }  \\
  & \mbox{ model  \cite{Buchmuller:1980su} } & \mbox{ model \cite{Bodwin:2006dm}} & \mbox{\cite{Bodwin:2006dn} } & \mbox{sum rules} \\
\hline
n=2  & 0.16  & 0.16  &  0.22 \pm 0.14   &  0.18~^{+0.05}_{-0.07}  \\
\hline
n=4  & 0.042 & 0.046 &   0.085 \pm 0.110   & 0.051~^{+0.031}_{-0.031}  \\
\hline
n=6  & 0.015 & 0.016 &  0.039 \pm 0.077  & 0.017~^{+0.016}_{-0.014}  \\
\hline
\end{array}$$
\caption{The moments of LCWF obtained within different approaches. In the second and third columns the moments calculated in the framework of 
Buchmuller-Tye and Cornell potential models are presented. In the fourth column NRQCD predictions for the moments are presented. 
In last column the results obtained within QCD sum rules are shown. }
\end{table}

It is not difficult to understand why these additional extremums appear. To do this 
let us differentiate equation ({\ref{p2}}) over $\xi$
\beq
\Phi' (\xi, \mu )= \frac { 2 \xi } { (1- \xi^2)^2 } \biggl [ 
(M_c^2+\mu^2)~ \psi \biggl (   \frac {\xi^2 M_c^2 + \mu^2} {1-\xi^2}  \biggr ) -
M_c^2~ \psi \biggl (   \frac {\xi^2 M_c^2 } {1-\xi^2}  \biggr )
\biggr ].
\label{def1}
\eeq
Equation (\ref{def1}) can be simplified if one recalls that the scale $\mu$ is 
much greater than characteristic momentum of relative motion of quark-antiquark pair inside 
the meson. This means that the function $\psi( (\xi^2 M_c^2 + \mu^2)/ (1-\xi^2) )$ in the first term is much less 
than $\psi( (\xi^2 M_c^2 )/ (1-\xi^2) )$ in the second term of equation (\ref{def1}) for not too large $\xi$. So, the 
first term gives small correction to the second and can be omitted to the first approximation.
Then equation (\ref{def1}) can be written as 
\beq
\Phi' (\xi, \mu )=- M_c^2 \frac { 2 \xi } { (1- \xi^2)^2 } 
~ \psi \biggl (   \frac {\xi^2 M_c^2 } {1-\xi^2}  \biggr ).
\eeq
It is well known that equal time wave function $\psi({\bf k^2})$ of 2$S$ state has one zero at some point ${\bf k_0^2}$. 
So it is clear that the function $\Phi' (\xi, \mu )$ changes sign at the points 
$\xi=\pm \sqrt { {\bf k_0^2} /( { \bf k_0^2}+ M_c^2 )}$ what corresponds to the two extremums of the 
function $\Phi (\xi, \mu )$. Moreover, the function $\Phi' (\xi, \mu )$ changes sign at $\xi=0$ what 
corresponds to the extremum at $\xi=0$. Obviously, if one regards the first term in 
equation (\ref{def1}) this will just shift the position of extremums. 

Applying the same arguments it is not difficult to prove the following statement: {\bf leading twist LCWF 
of $n$S state has $2 n+1$ extremums}. It should be noted here that our arguments are based on 
the relation between LCWF of leading twist and equal time wave function (\ref{p2}). In papers
\cite{Bodwin:2006dm, Ebert:2006xq} the other relations were proposed . Nevertheless, the statement written 
above remains true since it is valid for all relations of the type (\ref{p2}) where the 
function $\psi (t)$ can be represented as a product of equal time wave function and some function
$\chi (\xi) \sim 1 + O(v^2)$ for $\xi \sim v$.  

\section{The moments in the framework of NRQCD.}

To calculate the moments of LCWFs at leading order approximation in relative velocity 
one can use the following formula \cite{Braguta:2006wr}:
\beq
\langle \xi^n \rangle &=&  \frac { \gamma^n } { n +1 },
\label{nr2}
\eeq
where the constant $\gamma$ can be related to the matrix element of NRQCD operator $\gamma^2=\langle v^2 \rangle$. 
The value of $\langle v^2 \rangle$ can be calculated using the approach proposed in \cite{Bodwin:2006dn}
\beq
\langle v^2 \rangle = 0.65 \pm 0.42.
\label{resnr}
\eeq
There are different sources of error to result (\ref{resnr}). However, 
the main source of error is relativistic corrections to formula (\ref{nr2}). In (\ref{resnr}) the size 
of these corrections was estimated as  $\sim (\langle  v^2 \rangle)^2$. It is interesting to note 
that the value (\ref{resnr}) obtained at leading order approximation in relative velocity is 
very close to that obtained at next to leading order approximation $\langle  v^2 \rangle = 0.67$ \cite{bodwin}.

Using (\ref{nr2}), (\ref{resnr}) one can easily calculate the values of the moments. The results of this 
calculation are presented in the fourth column of Table I. The central values of the moments 
were calculated according to formulas (\ref{nr2}). The errors of the calculation of the moment 
$\langle \xi^{2 k} \rangle$ were estimated as $\sim k \langle v^2 \rangle~ \times \langle \xi^{2 k} \rangle$.

It is seen from Table I that within the error NRQCD prediction for the second moment is in agreement 
with potential model estimation, but the central values are rather far from each other.
For higher moments the difference between central values obtained within these approaches 
becomes more dramatic and the errors of the calculation within NRQCD are very large. 
From this one can draw a conclusion: although NRQCD can be applied to the calculation 
of the second moment of $2S$ state mesons, the predictions obtained within this approach 
for higher moments become unreliable due to large relativistic corrections.

It should be noted here that formula (\ref{nr2}) is very simple. So it is not difficult to guess 
that this dependence can be reproduced by the following 
function
\beq
\phi(\xi) = \frac 1 {2 \gamma} \theta (\gamma - |\xi| ),
\label{fnr}
\eeq
where $\theta(x)$ is the Heaviside step function. Function (\ref{fnr}) can be considered as the NRQCD LCWF obtained 
at leading order in relative velocity. If the velocity of quark-anitquark pair 
is infinitely small ($\gamma \to 0$) than $\phi(\xi)$ tends to $\delta(\xi)$ as it should be.

Function (\ref{fnr}) is very simple and 
it does not reproduce peculiarities of mesons. For instance, the only distinction of LCWFs of 
$1S$ and $2S$ states is different constants $\gamma$ of these mesons. However, from 
consideration of previous section it is known that the forms of these LCWFs are rather different. 
Actually, this is not surprising if one recalls that within NRQCD all  mesons with the same 
quantum numbers are described identically by one set of constants
and the peculiarities of each meson are contained in the values of these constants. At leading 
order approximation in relative velocity there is only one constant $\langle v^2 \rangle$. So 
it is not possible to reproduce peculiarities of each meson by the only constant. Probably, 
if one regards relativistic corrections and QCD radiative corrections to the expressions (\ref{nr2}) 
some properties will be restored.

\begin{figure}[t]
\begin{center}
\hspace*{-1cm}
\includegraphics[scale=0.89]{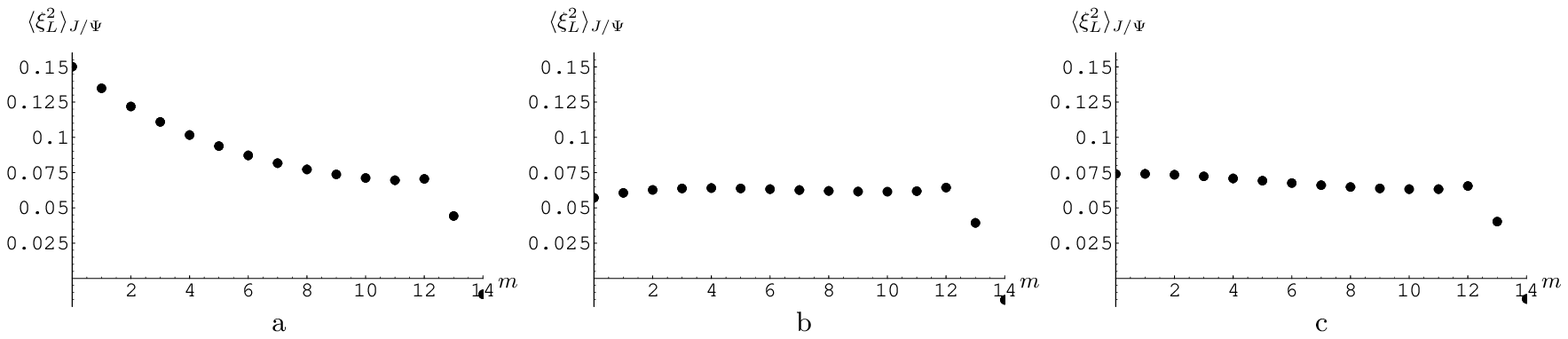} 
\caption{Sum rules for $\langle \xi^n_L \rangle_{J/\Psi}$ with different values of parameter $\langle \xi^n_L \rangle$: 
{\bf fig. a}~  $\langle \xi^n_L \rangle$=0; {\bf fig.  b}~ $\langle \xi^n_L \rangle$=0.22; {\bf fig. c}~ $\langle \xi^n_L \rangle$=0.18. }
\end{center}
\end{figure}

\section{The moments in the framework of QCD sum rules.}

\subsection{The moments of $\phi_L(\xi, \mu)$. }

In this section QCD sum rules \cite{Shifman:1978bx, Shifman:1978by} will be applied to the calculation of 
the moments \cite{Chernyak:1983ej, chernyak} of LCWFs $\phi_{L} (\xi, \mu)$.
To do this let us consider two-point correlator:
\beq
\Pi_L (z,q, n) = i \int d^4 x e^{i q x} \langle 0| T J_0(x) J_n (0) |0 \rangle = (zq)^{n+2} \Pi_L (q^2, n), 
\label{corL} \\ \nonumber
J_0 (x) = \bar Q(x) \hat z  Q(x), ~~~ J_n(0) = \bar Q(0) \hat z (i z^{\rho} {\overset {\leftrightarrow} {D}_{\rho}} )^n  Q(0), ~~ z^2=0.
\eeq
Sum rules for this correlator can be written as follows:
\beq
\frac {(f_{L})_{J/\Psi}^2 \langle \xi^n_L \rangle_{J/\Psi}}  { (M_{J/\Psi}^2+Q^2)^{m+1} } + 
\frac { (f_{L})_{\psi'}^2 \langle \xi^n_L \rangle_{\psi'}}  { (M_{\psi'}^2+Q^2)^{m+1} } = 
\frac 1 {\pi} \int_{4 m_c^2}^{s_0} ds ~ \frac {\mbox{Im} \Pi_{\rm pert}(s, n)} {(s+Q^2)^{m+1} }  + \Pi^{(m)}_{\rm npert}(Q^2, n) = \Pi_L (Q^2, n),
\label{smL}
\eeq
where 
The expressions for the functions $\mbox{Im} \Pi_{\rm pert}(s, n)$ and $\Pi^{(m)}_{\rm npert}(Q^2, n)$ can be 
found in paper \cite{Braguta:2007fh}. $(f_{L})_{J/\Psi}$ and $(f_{L})_{\psi'}$ are leptonic constants of $J/\Psi$ and $\psi'$ meson, 
$\langle \xi^n_L \rangle_{J/\Psi}$ and $\langle \xi^n_L \rangle_{\psi'}$ are the n-th moment 
of $J/\Psi$ and $\psi'$ mesons' LCWFs. To remain the designations introduced earlier, below 
$f_{L}$ and $\langle \xi^n_L \rangle$ will be used instead of $(f_{L})_{\psi'}$ and $\langle \xi^n_L \rangle_{\psi'}$. 

Numerical analysis of QCD sum rules (\ref{smL}) will be done similar to the numerical analysis in paper \cite{Braguta:2006wr}. 
To weaken the role of unknown radiative corrections instead of sum rules (\ref{smL}) 
the ratio of sum rules with different $n$ will be considered:
\beq
\frac {  \langle \xi^n_L \rangle_{J/\Psi} + r~ \langle \xi^n_L \rangle a(m) }
{ 1 + r~ a(m) } = \frac { \Pi_L (Q^2, 0 ) } { \Pi_L (Q^2, n) },
\label{smmod}
\eeq
where $r = f_{L}^2/ (f_{L})_{J/\Psi}^2$, 
\beq
a(m)= \biggl ( \frac { M_{J/\Psi}^2+Q^2 } { M_{\psi'}^2+Q^2 } \biggr )^{m+1} ~~~~~~~
r=\frac {f_{L}^2} {(f_{L})_{J/\Psi}^2} = \frac {M_{\psi'} \Gamma (\psi' \to e^+ e^-) } {M_{J/\Psi} \Gamma (J/\Psi \to e^+ e^-)} 
\simeq 0.53
\eeq
To calculate the moments of LCWF $\phi_{L} (\xi, \mu)$ let us rewrite sum rules (\ref{smmod}) as
\beq
\langle \xi^n_L \rangle_{J/\Psi} = \frac { \Pi_L(Q^2, 0 ) } { \Pi_L(Q^2, n) } \bigl ( 1 + r~ a(m)  \bigr )
- \langle \xi^n_L \rangle ~r~a(m). 
\label{expr}
\eeq
First sum rules (\ref{expr}) for $n=2$ will be considered. To the first approximation let us disregard the contribution of $\psi'$ meson in the 
right hand side of equation (\ref{expr}), as it was done in paper \cite{Braguta:2007fh} and take the value of 
the threshold $s_0$ equal to the 
threshold of $D$-mesons production $\sqrt s_0 \simeq 3.7$~GeV. The left hand side of equation (\ref{expr}) 
does not depend on $m$. The right hand side of (\ref{expr}) is a function of $m$. This function is plotted in 
Fig. 2a. It is seen that for too small values of $m$ ($m<10$) right hand side of equation (\ref{expr}) 
varies rather rapidly. This happens since there are large contributions from  higher resonances  
disregarded in model of physical spectral density  what invalidates sum rules (\ref{smL}), (\ref{expr}). 
Although for $m \gg m_1$ these contributions are strongly suppressed, it is not possible to apply sum rules for too 
large $m$ ($m>12$) since the contribution arising from higher dimensional vacuum condensates rapidly grows with 
$m$(see Fig. 2a) what also invalidates sum rules. It is seen from Fig. 2a that in the region $[10,12]$ left 
hand side of equation (\ref{expr}) $m$ varies very slowly. This is the region of applicability of sum 
rules (\ref{smL}), (\ref{expr}) where the resonance and the higher dimensional vacuum condensates contributions 
are not too large. Within the region of applicability the approximation of physical spectral density 
and the approximation of the contribution of vacuum condensates are valid and one can determine the value of the constant
$\langle \xi^2_L \rangle_{J/\Psi}$. Thus one gets
\beq
\langle \xi^2_L \rangle_{J/\Psi} = 0.07. 
\eeq
This value coincides with that  found in paper \cite{Braguta:2007fh}. 

As it was noted above due to the contribution of higher resonances sum rules (\ref{expr}) is spoiled 
in low $m$ region. Evidently, the inclusion one resonance succeeding  $J/\Psi$-meson will improve 
sum rules (\ref{expr}) in the region of low $m$. The parameter $\langle \xi^2_L \rangle$ can
be chosen so that to attain best fit of right hand side of equation (\ref{expr}) to the constant  
$\langle \xi^2_L \rangle_{J/\Psi}$. The calculation shows that the best fit can be obtained if 
$\langle \xi^2_L \rangle=0.22$. Right hand side of sum rules (\ref{expr}) at 
$\langle \xi^2_L \rangle=0.22$ as a function of $m$ is shown in Fig. 2b. 

From Fig. 2b it is seen that if $\psi'$ meson with $\langle \xi^2_L \rangle=0.22$ is included into the sum rules,  
the agreement between right and left hand sides of equation (\ref{expr}) becomes much better. 
From Fig. 2b one also sees that in the region $m \in [0,4]$ 
right hand side of sum rules (\ref{expr}) is rising function of $m$. This seems rather 
strange since if one includes charmonium meson succeeding $\psi'$ meson to sum rules, 
right hand side of equation (\ref{expr}) will become decreasing function of $m$. Perhaps,
this strange behavior originates from the following fact. In the region of too low $m$ there are large contributions 
coming from higher resonances not included into physical spectral density. So, if one tries 
to regard these contributions by the only resonance -- $\psi'$ meson, this will lead to an 
overestimation of the value of $\langle \xi^2_L \rangle$. This problem can be partially
removed if, in addition to the requirement to achieve the best fit of both sides of sum rules, 
the following requirement will be imposed: right hand side of equation (\ref{expr}) 
must be decreasing function of $m$. Thus one gets $\langle \xi^2_L \rangle=0.18$.
The right hand side of sum rules (\ref{expr}) as a function of $m$ with $\langle \xi_L^2 \rangle=0.18$ 
is plotted in Fig. 2c.

There are many sources of uncertainty of the calculation fulfilled above. The first one appears 
due to the uncertainty in sum rules parameters $m_c$ and $\langle {\alpha_s} G^2 /\pi \rangle$ 
\cite{Braguta:2007fh}. The calculation shows that the uncertainties due to the variation 
of $m_c$ and $\langle {\alpha_s} G^2 /\pi \rangle$ are not very important (not greater than 10\%). 
For this reason this source of uncertainty 
will not be considered in the calculation. Probably, the unknown contribution of QCD radiative corrections 
to the spectral density is much more important, but it is difficult to estimate its contribution. 
Another very important source of uncertainty results from the unknown value of the threshold
parameter $\sqrt s_0$. This parameter determines the energy from which continium contribution 
to sum rules appears. It is difficult to calculate the value of $s_0$, 
one can only claim that it is not very far from the threshold of $D$-mesons production 
$\sqrt s_0 \simeq 3.7$~GeV. In the calculation carried out in this paper it will be assumed
that $\sqrt s_0$ belongs to the interval $3.7 \pm 0.5$ GeV. The interval chosen in such a way 
is rather broad and it contains all intervals common for QCD sum rules analysis.
It should be noted here that the error due to the variation of $s_0$ within this interval is rather large and
below it will be considered as the error of the calculation. 

Applying the method discussed above for higher moments one gets the results:
\beq
\langle \xi^2_L \rangle &=& 0.18~^{+0.05}_{-0.07}, \nonumber \\ 
\langle \xi^4_L \rangle &=& 0.051~^{+0.031}_{-0.031}, \nonumber \\ 
\langle \xi^6_L \rangle &=& 0.017~^{+0.016}_{-0.014}.
\label{res}
\eeq
The central values of the moments have been calculated at $\sqrt s =3.7$ GeV. The errors of the 
calculation appears due to the variation of the threshold parameter $\sqrt s_0$ within the interval 
$3.7 \pm 0.5$ GeV. Physically this variation can be considered as a simulation of the contributions 
of higher charmonium states and continuum to the moments of LCWF $\phi_L (\xi, \mu)$. 
From this perspective the error of the calculation is rather large since the contributions 
from $\psi'$ meson, higher resonances and continuum are not well separated in sum rules (\ref{corL}). 
All these contributions appear approximately at $\sqrt s = 3.7$ GeV. So one can conclude that,
although this source of uncertainty can be diminished, it will remain to be the main source of uncertainty of the calculation.
From results (\ref{res}) one sees that the error of the calculation rises as number of the moment increases. 
Evidently, this happens since the larger the number of the moment the larger the sensitivity of this moment
to higher charmonium states and continuum. 

Results of the calculation (\ref{res}) are presented in the fifth column of Table I. It is seen
from this table that, although the accuracy of the results obtained within sum rules is better 
than NRQCD predictions for the moments, the error of the calculation is still rather large. 
It should be noted also that QCD sum rules predictions for the moments are in better 
agreement with potential models than with NRQCD results. The central values of NRQCD predictions
seems to be overestimated. 

\begin{figure}[t]
\begin{center}
\hspace*{-1cm}
\includegraphics[scale=0.89]{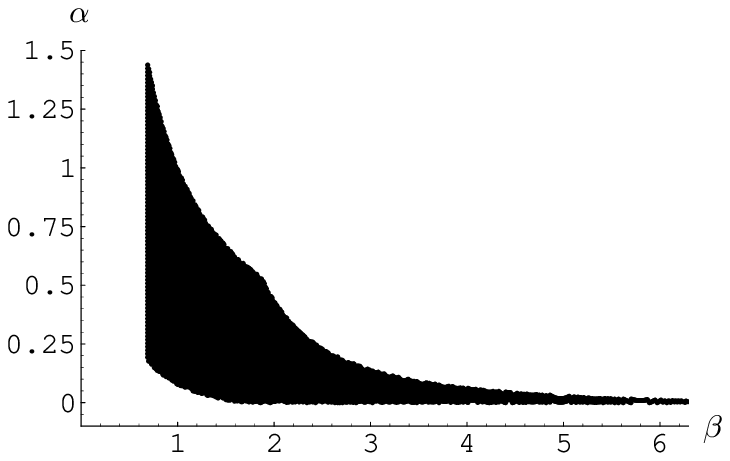} 
\caption{ Allowed region for parameters $(\beta, \alpha)$ ( model (\ref{model}) ) is painted black.}
\end{center}
\end{figure}

\subsection{The moments of $\phi_T(\xi, \mu)$ and $\phi_{\eta_c}(\xi, \mu)$. }

It is not difficult to derive sum rules for $\phi_T(\xi, \mu)$ and $\phi_{\eta_c}(\xi, \mu)$.
For instance, to calculate the moments of $\phi_{\eta_c}(\xi, \mu)$ one should consider two-point 
correlator:
\beq
\Pi_{\eta} (z,q, n) = i \int d^4 x e^{i q x} \langle 0| T J_0(x) J_n (0) |0 \rangle = (zq)^{n+2} \Pi_{\eta} (q^2, n), 
\nonumber \\ 
J_0 (x) = \bar Q(x) \gamma_5 \hat z  Q(x), ~~~ J_n(0) = \bar Q(0) \gamma_5 \hat z (i z^{\rho} {\overset {\leftrightarrow} {D}_{\rho}} )^n  Q(0), ~~ z^2=0.
\label{coreta}
\eeq
Sum rules for this correlator can be written as 
\beq
\frac {f_{\eta_c}^2 \langle \xi^n \rangle_{\eta_c}}  { (M_{\eta_c}^2+Q^2)^{m+1} } + 
\frac {f_{\chi_{c1}}^2 \langle \xi^n \rangle_{\chi_{c1}}}  { (M_{\chi_{c1}}^2+Q^2)^{m+1} } + 
\frac {f_{\eta_c'}^2 \langle \xi^n \rangle_{\eta_c'}}  { (M_{\eta_c'}^2+Q^2)^{m+1} } = 
\frac 1 {\pi} \int_{4 m_c^2}^{s_0} ds ~ \frac {\mbox{Im} \Pi_{\rm pert}(s, n)} {(s+Q^2)^{m+1} }  + \Pi^{(m)}_{\rm npert}(Q^2, n),
\label{smeta}
\eeq
where  $\langle \xi^n \rangle_{\eta_c}$, $\langle \xi^n \rangle_{\chi_{c1} }$ and 
$\langle \xi^n \rangle_{\eta_c'}$ are moments of leading twist LCWF of $\eta_c, \chi_{c1}, \eta_c'$ mesons,  
the constants $f_{\eta_c}, f_{\chi_{c1}}, f_{\eta_c'}$ are defined as
\beq
\langle 0| {\bar Q} (0) \gamma_{\alpha} \gamma_5 Q(0) |M(p) \rangle &=& i
f_{M} p_{\alpha}, ~~~~ M=\eta_c, \chi_{c1}, \eta_c'.
\eeq
One sees that in addition to $\eta_c'$ meson there is contribution of $\chi_{c1}$ meson. 
Since $\chi_{c1}$ meson is $P$ wave meson, its contribution is a little bit suppressed. 
Nevertheless, sum rules (\ref{smeta}) has one additional unknown parameter $\langle \xi^n \rangle_{\chi_{c1} }$ 
and this leads to worsening of sum rules predictions in comparison to case considered above. 
Similar situation takes place for $\phi_T(\xi, \mu)$, where there is contribution of $h_c$ charmonium meson.

From this one can conclude that, unfortunately, QCD sum rules cannot distinguish LCWF $\phi_L(\xi, \mu)$, $\phi_T(\xi, \mu)$ 
and $\phi_{\eta_c}(\xi, \mu)$ and it is not possible to calculate 
the moments of $\phi_T(\xi, \mu)$ and $\phi_{\eta_c}(\xi, \mu)$ with the accuracy better than the accuracy of 
the moments $\langle \xi^2_L \rangle_{J/\Psi}$. This makes the calculation of the moments of $\phi_T(\xi, \mu)$ 
$\phi_{\eta_c}(\xi, \mu)$ within QCD sum rules rather pointless. Below hypothesis 
(\ref{hy}) with moments (\ref{res}) will be used.

\section{The model for the functions $\phi_{\eta, L, T} (x, \mu)$.}

\begin{figure}[t]
\begin{center}
\hspace*{-1cm}
\includegraphics[scale=0.89]{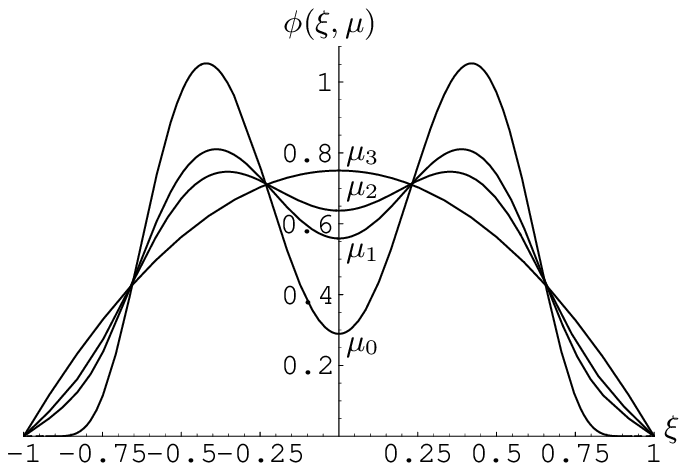} 
\caption{ The LCWF (\ref{model}) at scales $\mu_0 =1.2~\mbox{GeV} , \mu_1 = 10 ~\mbox{GeV}, \mu_2 =100 ~\mbox{GeV}, \mu_3 = \infty$.}
\end{center}
\end{figure}

Unfortunately, the methods applied in this paper to the calculation of the moments 
do not allow one to distinguish LCWFs $\phi_{\eta, L, T} (x, \mu)$. For this reason, below 
these functions are assumed to be equal to some function $\phi (x, \mu)$ at scale $\mu \sim m_c$. This section 
is devoted to the construction of the model for this function based on the results 
obtained within QCD sum rules. Results (\ref{res}) are defined at scale $\mu \sim m_c$ \cite{Braguta:2006wr}. 
In the calculations it will be assumed that these results are defined at  scale $\mu_0=1.2$ GeV$ \sim m_c$.

In papers \cite{Braguta:2006wr, Braguta:2007fh}  it was proposed one parametric model of LCWFs 
of $\eta_c$ and $J/\Psi$ mesons at scale $\mu_0=1.2$ GeV. To reproduce the results 
obtained in this paper this function can be modified by additional factor $(\alpha+\xi^2)$
\beq
\phi(\xi, \mu=\mu_0) = c(\alpha, \beta ) (1- \xi^2) (\alpha+\xi^2) \mbox{exp} \biggl (  -  \frac {\beta} {1-\xi^2} \biggr )
= c(\alpha, \beta ) (1- \xi^2) \Phi(\xi, \mu=\mu_0),
\label{model}
\eeq
Potential model calculation of $\Phi(\xi, \mu \sim m_c)$ tells us that this function  is positive 
and it has three extremums. Below it will be assumed that  these properties remain true for real 
function $\Phi(\xi, \mu=\mu_0)$. To meet the first requirement one can suppose that $\alpha \geq 0$. It is not difficult 
to show that the function $\Phi(\xi, \mu \sim m_c)$ has extremums at 
$\xi=0, \xi^2=(2+ \alpha - \sqrt {(2+\alpha^2)-4 (1-\alpha \beta)} )/2$. 
Two additional extremums of the function $\Phi(\xi, \mu \sim m_c)$ are beyond the physical region $\xi \in [-1,1]$. So, to meet the second requirement
- the function $\Phi(\xi, \mu=\mu_0)$ must have three extremums - one should impose the condition $\alpha \beta<1$.

Further let us find the region where the constant $\beta$ can vary. This can be done in the framework of Borel version 
of QCD sum rules \cite{Reinders:1984sr} where this constant can be expressed 
through the Borel parameter $M$ as follows $\beta=4 m_c^2/M^2$. The value of Borel parameter cannot 
be too small ($M>1$ GeV), otherwise the vacuum condensates contributions become too large. At the same time 
Borel parameter cannot be too large ($M<3$ GeV) otherwise the contributions of higher resonances become too large.
Thus one gets the assessment of the interval where the constant $\beta$ can vary $\beta \in (0.69, 6.25)$.
Now it causes no difficulties to find allowed region of the constants $\alpha, \beta$. This region
is painted black in Fig. 3. 

The central values of the second and the forth moment can be obtained within model (\ref{model}) 
if the values of the constants $(\alpha, \beta)$ are equal to $(0.027, 2.49)$. If one fixes the 
value of the constant $\alpha=0.027$ than, to attain the agreement of the model (\ref{model}) with 
the results (\ref{res}) for the second moment, the constant $\beta$ can vary within the interval $\beta \in (1.4 ,5.7)$.
Similarly if the constant $\beta$ is fixed at $2.49$ than the constant $\alpha$ can vary within the interval
$\alpha \in (0 , 0.35)$. 

Now let us consider model (\ref{model}) with the central values $\alpha=0.027, \beta=2.49$. As 
it was noted above model (\ref{model}) with these values of the constants $\alpha, \beta$ 
is defined at scale $\mu=\mu_0$. It is not difficult to 
calculate this function at any scale $\mu > \mu_0$ using conformal expansion \cite{Chernyak:1983ej}. 
This calculation will be done only for the function $\phi_{L} (x, \mu)$. 
The function $\phi_{L} (x, \mu)$ at scales $\mu_0 =1.2~\mbox{GeV} , \mu_1 = 10 ~\mbox{GeV}, 
\mu_2 =100 ~\mbox{GeV}, \mu_3 = \infty$
are shown in Fig. 4. The moments of this LCWF at scales $\mu_0 = 1.2~\mbox{GeV}, \mu_1 = 10 ~\mbox{GeV}, \mu_2 =100 ~\mbox{GeV}, \mu_3 = \infty$
are presented in second, third, fourth and fifth columns of Table II.  

In papers \cite{Braguta:2006wr, Braguta:2007fh} it was shown that due to evolution 
LCWFs of $1S$ state have some interesting properties: the violation of nonrelativistic QCD velocity scaling rules, 
appearance of relativistic tail and improvement of the accuracy of the model. LCWFs of $2S$ states 
have similar properties and in this paper these properties will not
considered.

Now let us consider two different models (\ref{model}): Model I $(\alpha=0, \beta=2.5)$ and Model II $(\alpha=0.2, \beta=2.5)$.
LCWF $\phi(\xi, \mu=\mu_0)$ of these models are shown in Fig.5a. LCWF of Model I has the following moments 
$\langle \xi^2 \rangle=0.21, \langle \xi^4 \rangle=0.061$, Model II has the moments $\langle \xi^2 \rangle=0.12, \langle \xi^4 \rangle=0.031$. 
It is seen that Model I is considerably wider than  Model II. In addition,  Models I and II are physically different.
Really, suppose the meson with momentum $p$ has LCWF of Model I. It is seen from Fig. 5a that this LCWF has rather sharp 
extremums at $|\xi| \sim 0.5$. This means that within this model it is not possible to produce $2S$ state 
charmonium meson from quark-antiquark pair with small relative momentum. Contrary to Model I, within Model II 
it is possible for quark-antiquar pair to have small relative momentum. Unfortunately, the uncertainties of 
results (\ref{res}) are rather large. So, both models are allowed. One can only assert that the model of 
LCWF with central values of parameters $\alpha=0.027, \beta=2.49$ is very similar to Model I. 
In addition, the forms of LCWF obtained within potential models (see Fig. 1) are similar to Model I.
It should be noted here that at leading order approximation of NRQCD quark-antiquark pair has zero relative momentum. 
So this approximation is in contradiction with Model I. 

The effect considered above takes place at scale $\mu=\mu_0$. To understand what happens at larger scales 
one should evolve Models I and II from scale $\mu_0$ to larger scales. LCWFs of Models I and II at scale $\mu=10$ GeV
are shown in Fig.5b. It is seen from this plot that the effect is not so dramatic as it is at scale $\mu_0$. 
This result is in agreement with the property of LCWFs discussed above: the larger the scale the less difference between
different models of LCWF.

\begin{table}
$$\begin{array}{|c|c|c|c|c|}
\hline \langle \xi^n \rangle & \phi(\xi, \mu_0=1.2~\mbox{GeV}) & \phi(\xi, \mu_1=10~\mbox{GeV}) & \phi(\xi, \mu_2=100~\mbox{GeV})
& \phi(\xi, \mu_3=\infty )  \\
  \hline
n=2  &  0.18 & 0.19  & 0.19 &  0.20 \\
\hline
n=4  & 0.051 &  0.068 & 0.074 & 0.086  \\
\hline
n=6  & 0.018 & 0.032 & 0.037 & 0.048  \\
\hline
\end{array}$$
\caption{ The moments of LCWF (\ref{model}) proposed in this paper at scales $\mu_0 = 1.2~\mbox{GeV}, \mu_1 = 10 ~\mbox{GeV}, \mu_2 =100 ~\mbox{GeV}, \mu_3 = \infty$
are presented in second, third, fourth and fifth columns.}
\end{table}

\section{Conclusion}

In this paper the moments of leading twist light cone wave functions (LCWF) of $2S$ state 
charmonium mesons have been calculated within three approaches. 
In the first approach Buchmuller-Tye and Cornell potential models were applied 
to the calculation of the moments of LCWFs. 
In the second approach the moments of LCWFs were calculated in the 
framework of NRQCD. In the third approach the method QCD sum rules was applied to the calculation of the moments. 
Although, the results of the calculation are in reasonable agreement with each other, the errors of 
the calculation are rather large. As the result, it is not possible to distinguish different LCWFs form each other. 

Similarly to the study of LCWFs of $1S$ state charmonium mesons \cite{Braguta:2006wr, Braguta:2007fh}, the most 
accurate results were obtained within QCD sum rules. 
Using these results two parametric model of LCWFs of $2S$ states was proposed. This model can be used 
in the calculation of different hard exclusive processes with $2S$ charmonium mesons production.

\begin{acknowledgments}
The author thanks A.K. Likhoded, V.V. Kiselev and A.V. Luchinsky for useful discussion and help in preparing this paper.
The author thanks G.T. Bodwin for useful discussion.
This work was partially supported by Russian Foundation of Basic Research under grant 07-02-00417, Russian Education
Ministry grant RNP-2.2.2.3.6646, CRDF grant Y3-P-11-05, president grant MK-2996.2007.2 and the Dynasty foundation.
\end{acknowledgments}

\begin{figure}[t]
\begin{center}
\hspace*{-2.cm}
\includegraphics[scale=0.9]{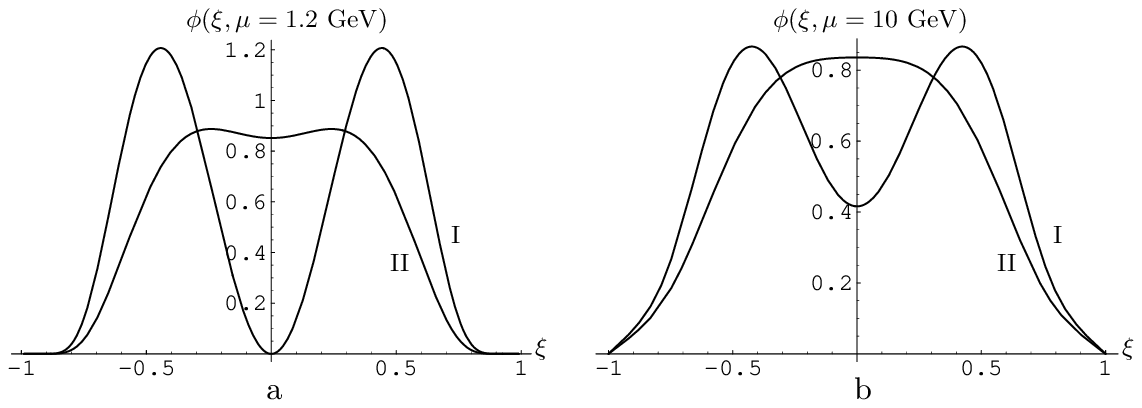} 
\caption{ LCWFs (\ref{model}) at scales: {\bf fig. a}~ $\mu=1.2$ GeV; {\bf fig. b}~ $\mu=10$ GeV with different parameters: 
Model I $(\alpha=0, \beta=2.5)$ and Model II $(\alpha=0.2, \beta=2.5)$.}
\end{center}
\end{figure}


\begin{thebibliography}{**}


\bibitem{Chernyak:1983ej}
  V.~L.~Chernyak and A.~R.~Zhitnitsky,
  Phys.\ Rept.\  {\bf 112}, 173 (1984).


\bibitem{Bodwin:1994jh}
  G.~T.~Bodwin, E.~Braaten and G.~P.~Lepage,
  Phys.\ Rev.\ D {\bf 51}, 1125 (1995)
  [Erratum-ibid.\ D {\bf 55}, 5853 (1997)]
  [arXiv:hep-ph/9407339].
  
  
  
\bibitem{Braaten:2002fi}
  E.~Braaten and J.~Lee,
  Phys.\ Rev.\ D {\bf 67}, 054007 (2003)
  [Erratum-ibid.\ D {\bf 72}, 099901 (2005)]
  [arXiv:hep-ph/0211085].




\bibitem{Liu:2002wq}
  K.~Y.~Liu, Z.~G.~He and K.~T.~Chao,
  Phys.\ Lett.\ B {\bf 557}, 45 (2003)
  [arXiv:hep-ph/0211181].

\bibitem{Liu:2004ga}
  K.~Y.~Liu, Z.~G.~He and K.~T.~Chao,
  arXiv:hep-ph/0408141.

\bibitem{Zhang:2005ch}
  Y.~J.~Zhang, Y.~j.~Gao and K.~T.~Chao,
  Phys.\ Rev.\ Lett.\  {\bf 96}, 092001 (2006)
  [arXiv:hep-ph/0506076].


\bibitem{Bodwin:2006ke}
  G.~T.~Bodwin, D.~Kang, T.~Kim, J.~Lee and C.~Yu,
  arXiv:hep-ph/0611002.
  
  \bibitem{He:2007te}
    Z.~G.~He, Y.~Fan and K.~T.~Chao,
    Phys.\ Rev.\  D {\bf 75}, 074011 (2007)
    [arXiv:hep-ph/0702239].
  

\bibitem{Bodwin:2006dm}
  G.~T.~Bodwin, D.~Kang and J.~Lee,
  Phys.\ Rev.\  D {\bf 74}, 114028 (2006)
  [arXiv:hep-ph/0603185].



\bibitem{Ebert:2006xq}
  D.~Ebert and A.~P.~Martynenko,
  Phys.\ Rev.\  D {\bf 74}, 054008 (2006)
  [arXiv:hep-ph/0605230].

  \bibitem{Ma:2006hc}
    J.~P.~Ma and Z.~G.~Si,
    arXiv:hep-ph/0608221.
  

\bibitem{Choi:2007ze}
  H.~M.~Choi and C.~R.~Ji,
  arXiv:0707.1173 [hep-ph].




\bibitem{Braguta:2006wr}
  V.~V.~Braguta, A.~K.~Likhoded and A.~V.~Luchinsky,
  Phys.\ Lett.\  B {\bf 646}, 80 (2007)
  [arXiv:hep-ph/0611021].




\bibitem{Braguta:2007fh}
  V.~V.~Braguta,
  Phys.\ Rev.\  D {\bf 75}, 094016 (2007)
  [arXiv:hep-ph/0701234].








  
  


\bibitem{Rosner:2001nm}
  J.~L.~Rosner,
  Phys.\ Rev.\  D {\bf 64}, 094002 (2001)
  [arXiv:hep-ph/0105327].
  


  \bibitem{Brodsky:1981jv}
    S.~J.~Brodsky, T.~Huang and G.~P.~Lepage,
    In *Banff 1981, Proceedings, Particles and Fields 2*, 143-199.

\bibitem{Buchmuller:1980su}
  W.~Buchmuller and S.~H.~H.~Tye,
  Phys.\ Rev.\ D {\bf 24}, 132 (1981).

\bibitem{Eichten:1978tg}
  E.~Eichten, K.~Gottfried, T.~Kinoshita, K.~D.~Lane and T.~M.~Yan,
  Phys.\ Rev.\ D {\bf 17}, 3090 (1978)
  [Erratum-ibid.\ D {\bf 21}, 313 (1980)].



  \bibitem{Bodwin:2006dn}
    G.~T.~Bodwin, D.~Kang and J.~Lee,
    Phys.\ Rev.\ D {\bf 74}, 014014 (2006)
    [arXiv:hep-ph/0603186].

\bibitem{bodwin} 
  G.~T.~Bodwin, private communication



\bibitem{Shifman:1978bx}
  M.~A.~Shifman, A.~I.~Vainshtein and V.~I.~Zakharov,
  Nucl.\ Phys.\ B {\bf 147}, 385 (1979).
  
\bibitem{Shifman:1978by}
  M.~A.~Shifman, A.~I.~Vainshtein and V.~I.~Zakharov,
  Nucl.\ Phys.\ B {\bf 147}, 448 (1979).


\bibitem{chernyak}
V.~L.~Chernyak and A.~R.~Zhitnitsky, 
  Nucl.\ Phys.\ B {\bf 201}, 492 (1982)
  








\bibitem{Reinders:1984sr}
  L.~J.~Reinders, H.~Rubinstein and S.~Yazaki,
  Phys.\ Rept.\  {\bf 127}, 1 (1985).

\end{thebibliography}
\end{document}